**Expertise Revisited I - Interactional expertise**

**Harry Collins and Robert Evans**

Centre for the Study of Knowledge Expertise and Science (KES)
School of Social Sciences
Cardiff University
Cardiff CF10 3WT
UK

Corresponding author
CollinsHM@cf.ac.uk
+44 (0)2921409637

**Abstract**

In Part I of this two part paper we try to set out the 'essence' of the notion of interactional expertise by starting with its origins.  In Part II we will look at the notion of contributory expertise.  The exercise has been triggered by recent discussion of these concepts in this journal by Plaisance and Kennedy and by Goddiksen.

**Keywords**

Interactional Expertise; Contributory Expertise; Technical Phase; Political Phase; Policy

**1.    Introduction: Four streams of interactional expertise**

Concepts that are widely taken up will be developed and adapted by others, the authors' only privilege being special access to the origins of the ideas.[1]  Triggered by some recent critiques and suggested modifications, in 'Part I' of this paper we try to set out the deep meaning of interactional expertise in so far as it can be distilled from its origins pointing out that the central idea is the separation of language and practice which, to give it fresh salience, we will call the 'separation principle'.  In 'Part II' we will look at what it means to make contributions to specialist debates.  Along the way we will introduce other new ideas including 'ubiquitous interactional expertise' and we will split the idea of referred expertise into two parts and also point to the meta-expertise element of interactional and contributory expertise.

Memory is unreliable so for the purpose of tracing origins we used a special utility ('astrogrep') to explore old computer archives, reminding ourselves of much that we had forgotten and coming up with a number of surprises. Thus, in many talks we have been describing the origins of the idea of interactional expertise as solely to do with fieldwork experience in the late 1990s but, though the fieldwork was the origin of the *term* 'interactional expertise', the *idea* goes back further to debates about artificial intelligence and the sociology of knowledge. We conclude that four channels feed the *idea* of interactional expertise, as shown in Figure 1. The backbone of the concept is a philosophical stream; this is shown in black with the first invocation of the idea being the second entry, dated 1995/6 and labelled as the separation principle. A 2004 article (Collins 2004b) in a philosophy journal unites the philosophy stream with an 'imitation game tributary' and a 'fieldwork tributary'. There is also a 'sociology and policy' channel which, on pain of some convoluted hydraulics, is mostly downstream from the others in a sideways kind of way.

---

[1] The papers and books introducing the concept of interactional expertise and other related expertises have been cited many more than 2,300 times and a Google search for 'interactional expertise' reveals very wide discussion of the idea.

| IMITATION GAME | PHILOSOPHY | FIELDWORK | SOCIOLOGY POLICY |
|---|---|---|---|
| colspan="4" | **Winch 1958 Wittgenstein 1953<br>The sociological interpretation of Wittgenstein**<br>*1974 1975 ...* | | |
| *1990*<br>*Artificial Experts*<br>Pre-IE experiments | **1995** *1996*<br>**SEPARATION PRINCIPLE**<br>Separate language and practice<br>Embodiment | **1997** *2002*<br>*The term 'interactional expertise'* | Epstein Sheep farmers |
| **2004** Colour blind, perfect pitch experiments begin | | **2002** *2004*<br>*Gravity's Shadow* | Imagined problem of GW and power lines!<br>Managers |
| colspan="3" | **2003?** *2004* **THE SYNTHESIS** | | Peer reviewers<br>Committees<br>Sports coaches etc |
| **2005** Collins as GW physicist<br>**2007** The blind<br>**2007** Classroom Experiments on Gays etc | **2006** *2007*<br>Strong Interactional Hypothesis | | **2007**<br>Crossing linguistic divides |
| **2011** Large Scale Experiments begin | | | *2011*<br>Special IE Division of Labour |
| | | *2013*<br>*Gravity's Ghost and Big Dog* | **2014** Ubiquitous IE Social 'glue'<br>**2014** The Owls |

*Figure 1: Evolution of Idea of interactional expertise (dates of publications in italics)*

After explaining the four-channel model of interactional expertise (sometimes referred to below as IE), we are going to suggest that certain critiques and proposed amendments are in danger of creating an 'interactional expertise lite' which would have very wide application but would lose sight of the force that drove the concept's development. 'Lite' terms are popular because they give the impression of resolving deep philosophical problems without the usual detailed work; instead mere use of the term is taken to be enough. Thus the invocation of 'actor network theory' in its lite incarnation allows social, political and material factors to be combined in whichever way the analyst prefers without worrying about which has causal priority in the real world, what the methodology is for establishing such things, or the theory's underlying and strange metaphysics of agency; the term 'trading zone' can be used in a lite way to resolve anxieties about the difficulty of communication across linguistic and cultural boundaries without looking deeply into the detail of how such things work; and the same often goes for the term 'boundary object', which is invoked to 'explain' how different communities collaborate despite different interests and objectives.[2] Even though it might widen its appeal still further, we think it could be prejudicial to the notion of expertise in general if the term 'interactional expertise' comes to be used to licence contributions to technical debates without the grounds of the expertise being established independently of the mere invocation of the term.

In Part II we will confront the fact that interactional expertise has had an ambivalent and unsatisfactory relationship with the idea of 'contributory expertise'. We will not be able to resolve all the problems but we will assemble our newly evolved understanding of what it is to make a contribution to specialist domains and to technological decision-making in the public domain. The immediate occasion for revisiting these concepts is the 2014 discussion of interactional expertise by Plaisance and Kennedy (2014) (PK) and the debate between Goddiksen (2014), and Reyes-Galindo and Duarte (2015 – RGD). We believe the new papers published here provide a clearer framework for their project.

---

[2] The term 'Latour Lite', whose origin is unknown (Collins thought Shapin invented it but he denies it), refers to the many purported uses of the work of Bruno Latour which do not get the heart of his position nor its implications. Latour (private communication with Collins, 1/09/2011) acknowledges the problem but says it is not his fault. For critiques of Actor Network Theory which are germane to the problem, see Collins and Yearley (1992) and Collins (2012). For an analysis which gives 'trading zones' and 'boundary objects' places in a more detailed scheme of cross-language communication see Collins et al (2007).

## 2. The philosophy stream

The philosophical *idea* of interactional expertise arose in the mid-1990s before the term was invented.[3] The first published appearance of *the term* 'interactional expertise' (IE) was in the 'Third Wave' paper by Collins and Evans, published in 2002 but this paper dealt with the concept in a superficial way. The first full discussion of the term is found in a 2004(b) paper entitled 'Interactional Expertise as a Third Kind of Knowledge'.

### *2.1 The sociological interpretation of Wittgenstein*

The more basic ideas that led to the concept can be dated from the much earlier interpretation of Wittgenstein that gave rise to the sociology of scientific knowledge as represented in the full-width black box at the head of Figure 1. This finds its most explicit formulation in David Bloor's (1983) argument that the later (e.g. 1953 / 1958) Wittgenstein is to be thought of as a sociologist as much as a philosopher.[4] In our case, the start was a reading of Winch's (1958 / 1988) *The Idea of a Social Science* – a book which presents itself as a philosophical critique of sociology. The reading of Winch with which we started 'stands Winch on his head' or, at least, 'on his side'. What is treated as crucial is Winch's argument that social and conceptual life are two sides of the same coin but the reading draws out the sociological implications of the idea instead of presenting it as a philosophical critique. In so far as a critique can be drawn from Winch's argument, philosophy is as much sociology as sociology is philosophy. This reading gives rise to what we will call 'the sociological interpretation Wittgenstein' and the consequent use of the work of the later Wittgenstein is similar to that recommended by Bloor. This interpretation of Wittgenstein has been accepted by only a minority of philosophers but has proved fruitful in the sociology of scientific knowledge and other domains. The Winch-inspired sociological interpretation of Wittgenstein gave the diverse streams of the idea of interactional expertise unity even though it was not fully expressed, at least not in print, until 2004.

---

[3] These earlier works are, in fact, referenced in Collins (2004b) but we had forgotten their relevance.

[4] See also Bloor (1976 / 1991)

## *2.2 The separation principle*

The sociological interpretation of Wittgenstein treats 'forms-of-life' in a Durkheimian way such that a social group's patterns of language and practice give rise to their understanding of the world.[5] The new thought that gives rise to interactional expertise is that the contributions to a form-of-life of language on the one hand, and practice on the other, can be analytically and empirically separated – this is what we are now calling the 'separation principle'. Under this model, language and practice *together* give rise to forms-of-life at the collective level but individuals can acquire a complete 'understanding' of a form-of-life through immersion in the *language alone* and without taking part in the practices. Here, 'understanding' implies, 'as tested in conversation or question-and-answer', not 'as demonstrated through practical accomplishment': practice cannot be learned from conversation, only an understanding of practice but, as will be argued, this understanding can be as good as practitioners' understanding when it comes to making technical judgements. Clearly, then, we must distinguish between *practicing* and *understanding practice*, with the latter being tested through linguistic accomplishment, not practical accomplishment. To repeat, this distinction works only for individuals: at the collective level languages are established through interaction with the practices pertaining to the community. The idea is well-explained in the following quotation which is probably its first expression in print.

---

[5] Which also gave rise to a very natural sociological/philosophical interpretation of the Kuhnian notion of paradigm (Kuhn 1996).

> Wittgenstein said that if a lion could speak we would not understand it. The reason we would not understand it is that the world of a talking lion - its `form of life' - would be different from ours. Bringing back Dreyfus's chair example, lions would not have chairs in their language in the way we do because lions' knees do not bend as ours do, nor do lions `write, go to conferences or give lectures'. ... But this does not mean that every entity that can recognise a chair has to be able to sit on one. That confuses the capabilities of an individual with the form of life of the social group in which that individual is embedded. Entities that can recognise chairs have only to share the form of life of those who can sit down. We would not understand what a talking lion said to us, not because it had a lion-like body, but because the large majority of its friends and acquaintances had lion-like bodies and lion-like interests. In principle, if one could find a lion cub that had the potential to have conversations, one could bring it up in human society to speak about chairs as we do in spite of its funny legs. It would learn to recognise chairs as it learned to speak our language. This is how the Madeleine case is to be understood [Madeleine is a woman, discussed by Oliver Sacks (1985 / 2011), who was almost totally disabled from birth yet, according to Sacks, was completely fluent in spoken language]; Madeleine has undergone linguistic socialization. In sum, the shape of the bodies of the members of a social collectivity and the situations in which they find themselves give rise to their form of life. Collectivities whose members have different bodies and encounter different situations develop different forms of life. But given the capacity for linguistic socialisation, an individual can come to share a form of life without having a body or the experience of physical situations which correspond to that form of life.[6]

Dreyfus is famous for claiming that computers cannot act 'intelligently' because they do not have bodies. The counter-example of Madeleine was put forward by AI-pioneer, Douglas Lenat, one of Dreyfus's critics, in order to show that one can learn language without having a human-like body and Collins agreed with Lenat on this point while still holding that computers cannot be intelligent because they cannot be socialised.[7] *The term* 'interactional expertise' was *not* used in these early contributions but the idea that language and practice must be separated if we are to be clear about the different relationship of the collectivity and the individual to a form-of-life is the key. This, of course, is the basis of IE. In turn, this depends on the idea that language is not formal, 'propositional knowledge', as it is treated by those keen to stress the practice element of forms-of-life, but is itself a tacit knowledge-laden form of social interaction (Collins, 2004b). Indeed, wherever readers see the word 'language' they should understand something like 'linguistic practice' (Collins, 2011) because language is far more than words and grammar.

---

[6] From a review of Hubert Dreyfus's book, *What Computers Still Can't Do*, (Collins 1996), which was published in the journal *Artificial Intelligence.*

[7] The argument continued in Collins's contribution to Hubert Dreyfus's *Festschrift* under the terms 'social embodiment thesis; 'individual embodiment thesis' and 'minimal embodiment thesis' (Collins 2000) and further published discussions followed in, for example, Selinger (2003), Collins (2004c) and Selinger, Dreyfus and Collins (2007). The philosophers (Dreyfus and Selinger) argued that the crucial point was that Madeleine had a body with a sense of front, back and so forth and that is why Lenat was wrong; the sociologist (Collins), argued that the interesting thing was the fluency that Madeleine could achieve with only a 'minimal body'. Selinger also instigated another approach and which gave rise to the 'Dispositions' line in the Periodic Table of Expertises (Collins and Evans 2007). In retrospect, we can call this approach the investigation of the social role of special interactional experts. Selinger and Mix (2004) argued that interactional experts might have special interactional skills (what we called 'interactive ability') in talking about expertises that were not their own and in translating between communities. Plaisance and Kennedy also pick up this idea in their critique (Plaisance and Kennedy 2014). This seems to us to be an entirely reasonable line to take but the sociology of interactional experts should not be confounded with the basic idea of IE: it draws on the basic idea but does not feed back into it.

Philosophically, interactional expertise as it first appeared makes a new claim about how forms-of-life work and how individuals fit into them: mostly people in a society come to understand each other through socialisation that involves sharing both practices and language but sometimes this understanding is via acquisition of the spoken language alone. It is argued that only in this way can we understand, for example, how those who are sufficiently physically challenged to be unable to engage in the common practices of their society can come to participate in normal conceptual life as a result of their immersion in the spoken discourse alone. We'll call this the 'linguistic definition' of IE.

### *2.3* *The strong interactional hypothesis*

The last entry in the philosophy stream is a bold conjecture Popper (1959 / 2002) – the 'strong interactional hypothesis' (SIH): in principle, the level of fluency in the language of a domain that can be attained by someone who is an interactional expert only is indistinguishable from that which can be attained by a full-blown contributory expert in any test involving language alone (e.g. Collins and Evans 2007:31)

The SIH stresses the philosophical nature of the concept: under the right circumstances an individual can *fully* understand the world through the medium of language alone. As a bold conjecture, the SIH may be wrong. The initial empirical foundation for it is Sacks's account of 'Madeleine' (see above) and this is not a reliable source because Sacks had a different purpose in mind. The SIH is meant to encourage more such experiments and observations – pressing thought and experiment to its limits is the job of a bold conjecture. In 2007, not really believing the SIH, we embarked on a series of imitation game tests on the blind in order to try to find the point at which it failed and were genuinely surprised that the limit did not appear even in such a hard case. The SIH could motivate more and better research. It is hard to prove or disprove the SIH because one must show that persons failing tests did have the opportunity to gain maximal interactional expertise, while it also has to be shown that persons passing the test had gained their understanding purely from linguistic sources. Both conditions are rare, hard to fulfil and very easy to forget.[8]

Societies, as opposed to individuals, cannot form and develop without both the language and the practices being constantly used and refreshed. This leads to the suggestion that IE on its own is parasitical in the sense that the 'practice language' spoken by an individual would atrophy, or become distorted, if not continually refreshed by contact with the talk of those who engage in the practices it encapsulates.[9] This is an idea that needs testing philosophically and practically, but it has already led to interesting questions about how IE is acquired, how much faster it can be acquired with exposure to practice and why, if this is so, it should be so.[10]

---

[8] We now think the SIH needs modification in the light of the effects *on early brain development* of certain congenital conditions.

[9] A practice language (Collins 2011), is the language associated with a particular set of practices, such as tennis-playing or doing gravitational wave physics. Goddiksen (2014) challenges the claim that practice languages are parasitical but not in a convincing way – see below.

## 3. The fieldwork tributary

The fieldwork tributary arose out of Collins's deep and extended immersion in the society of gravitational wave physicists which began in the mid-1990s. Towards the end of the 1990s he found himself involved in technical physics talk of the kind that physicists had with each other even though he was not a physicist. He called his ability to talk physics, 'interactional expertise' and the term was subsequently mentioned in print in the 'Third Wave paper' (Collins and Evans 2002).[11]

The first and most intensive discussion of how 'fieldwork interactional expertise' works is referred to on page 750 of *Gravity's Shadow* (Collins 2004a). It comprises an annotated transcript of a heated argument between Collins and Barry Barish, the Director of the LIGO project which took place in the year 2000. This transcript was cut from the published book on grounds of length. It can now be found at http://www.cardiff.ac.uk/socsi/kes/downloads under the heading, 'Annotated Interactional Expertise Debate'. Here is a pertinent section of the annotated transcript. The context is that Collins, basing his arguments on those of a disaffected LIGO scientist, Bob Spero, is claiming that Barish does not understand all that he needs to know about noise in big (4km) interferometers from the work that has been done on the prototype 40 meter interferometer. He is saying to Barish that he cannot build the big, $100M, machines with confidence before further fundamental research on noise is done. Barish is saying he does know enough or, at least, that Spero has not shown that he does not know enough. At a certain point Barish presents a potentially decisive argument. Collins's capitalised reflexive comment, written for publication, reveals the role of interactional expertise in fieldwork as it was being thought about at this juncture:

```
Barish:   Now, let me tell you one other thing.  This [sensitivity] is
[about a 10⁻¹⁹] meters displacement.  To get to LIGO sensitivity you
don't have to do [more than two] orders of magnitude better than this
[in noise reduction].  [If] You do this well [ie, as well as the 40
meter] over a longer baseline you gain because of the baseline.  This
is already at the level you have to do in LIGO, you just have to do it
over a longer baseline -- which is what I call engineering.  This
displacement sensitivity is what we asked for in LIGO -- [10⁻¹⁹] meters.
[[Collins: 40 compared to 4000 ...]]  Yeah we gain a factor of a
[hundred] giving 10⁻²¹ giving this curve for displacement sensitivity
and then we ... OK? {The 4 kilometre LIGO interferometers are designed
to have a displacement sensitivity of 10⁻²¹.}

      So this is already at the level [required in terms of noise
elimination] and there is no unknown source -- there's maybe a factor
of two, but nothing that is a factor of 20.

ASIDE TO THE READER -- READ THIS CAREFULLY BEFORE GOING ON.  IF YOU
HAVE UNDERSTOOD THE TECHNICAL PART OF THIS BOOK YOU SHOULD HAVE BEEN
ABLE TO FOLLOW THE ARGUMENT SO FAR.  LET ME REPEAT THE LAST POINT:
BARISH HAS SAID THAT THERE CANNOT BE ANY UNKNOWN NOISE SOURCES THAT
WILL PREVENT LIGO GETTING TO A SENSITIVITY OF 10⁻²¹ BECAUSE TO REACH
THAT SENSITIVITY ALL LIGO HAS TO DO IS BEAT DOWN THE NOISE TO THE SAME
LEVEL AS HAS ALREADY BEEN ACCOMPLISHED IN THE 40 METER; THE SHEER EXTRA
LENGTH OF LIGO WILL DO THE REST.

      IF YOU CAN UNDERSTAND THAT YOU MIGHT BE WONDERING WHAT I AM GOING
TO SAY TO BARISH NEXT BECAUSE, ON THE FACE OF IT, THIS LAST POINT IS A
`KILLER.'  IF ALL LIGO NEEDS TO DO TO REACH THE DESIGN SENSITIVITY IS
SCALE UP FROM THE 40 METER WHAT MORE RESEARCH WAS THERE FOR THE 40
METER TO DO?  HOW COULD THERE BE ANY ADDITIONAL FATAL NOISE SOURCES?
BOB SPERO AND THE 40 METER TEAM MUST SIMPLY HAVE BEEN WRONG AND MY
BELOVED METHODOLOGICAL RELATIVISM IS UNSUSTAINABLE -- AT LEAST BY ME.
SO IF YOU THINK YOU HAVE AS MUCH INTERACTIONAL EXPERTISE AS ME YOU
MIGHT BE ABLE TO WORK OUT WHAT I SHOULD SAY NEXT.  AND IF YOU CAN, IT
MAY BE THAT MY LONG SOJOURN WITH THE GRAVITATIONAL WAVE PHYSICISTS WAS
NOT REALLY NECESSARY.
```

> IF YOU CANNOT ANTICIPATE WHAT I WILL SAY NEXT, OR ANY EQUALLY SATISFACTORY ALTERNATIVE, IT SHOWS THAT YOU HAVE LESS INTERACTIONAL EXPERTISE THAN ME IN SPITE OF THE FEELING -- THAT I HOPE I HAVE ENGENDERED IN YOU BY MY CLEAR EXPOSITION OF THE SCIENCE IN THIS BOOK -- THAT YOU ACTUALLY KNOW AS MUCH AS ME. SO HAVE A THINK ABOUT IT AND SEE IF YOU CAN SEE HOW ONE MIGHT CONTINUE THE ARGUMENT.[12]

---

[10] Thus Ribeiro seems to have shown that those who have a deeper 'level of immersion' in practical matters tend to understand things better but one would hardly expect it to be otherwise, not least because the deeper your level of immersion in practice the more you are likely to be immersed in the corresponding linguistic discourse. So, in contrast to the way he presents his findings, they bear weakly, at best, on the SIH (Collins 2013b; Ribeiro 2013). Given such evidence as we have, it would be better if such research projects were driven by the idea that physical immersion in practice is *the condition for* immersion in the practice language with the responsibility being to prove this is wrong (Collins 2011).

[11] Incidentally, the term 'third wave' first appears in a draft by Collins written between November 1999 and June 2000 which refers to discussions already taking place at the Cardiff Centre for the Study of Knowledge, Expertise and Science (KES).

Collins: What configuration was this interferometer running in? Was it power-recycled at the time?

Barish: No.

Collins: So if you say to get to $10^{-21}$ you only need the extra length, why do you need to use power-recycling? {Here I am asking Barish whether it is not more than a matter of just scaling up.}

Barish: In order to get the shot noise limit, which is this for us [indicating figure], we need more light in the big interferometer. It's photons per frequency bin, or per second, so to fill up the 4 kilometre you need more light. {This is a technically important point because to reach the sensitivity required, the 4 kilometre interferometer is actually going to have to run in a configuration of the kind that was not fully explored by the 40 meter. This is the `power-recycled' configuration. I go on to exploit this point.}

Collins: So could there be some source of noise that's configuration dependent?

Barish: Of course.

Collins: Could it be that when you power-recycle [the interferometer] there could be some sorts of noise that aren't represented on this curve?

Barish: But [that's not] fundamental noise. Its noise that has to do with how well we bounce the light around or make a resonance, it's what I call technical noise. I agree there can be technical noise. ... Technical noise is boring, it's hard work, and they weren't very good at it on the 40 meter. Fundamental noise is a different matter.

The argument continued (as can be seen at http://www.cardiff.ac.uk/socsi/kes/downloads)

with Collins telling Barish that the previous director of LIGO, Robbie Vogt, had claimed that unknown and unanticipated noise sources appeared whenever the interferometer configuration was changed and Barish claiming that these should be seen as 'technical' and not 'fundamental'. As can be seen, interactional expertise as Collins experienced it in its fieldwork incarnation, is demanding and not easy to aquire.[13]

---

[12] This notion that a clear exposition of science gives rise to the feeling that you understand it while, in actuality, you don't, will prove important in our discussions of 'lay expertise' and bears on the 'paradigm cases' of interactional expertise that we will discuss below.

The fieldwork and artificial intelligence streams were treated independently – at least in published sources – until the connection was made in the 2004(b) paper.[14] It is, however, clear that both streams of thinking are related to the same new philosophical point: that a conceptual grasp of socially embedded practices can be acquired from spoken discourse alone. In the first stream this is exemplified by the situation of the congenitally disabled; in the second stream it is exemplified by the fieldworker immersed in an initially unfamiliar technical domain.

**4.    The imitation game tributary**

> A Turing-type test is a good test for computing ability precisely because it is a test of the extent to which a machine can be located in an interactional network without strain. Instead of asking about the innate ability of the machine one looks at its interactive competence; this is how we judge other things that interact with us. ... [they] fit neatly into our socio-cognitive networks (Collins 1990:190). [The other things being referred to are humans.]

---

[13] The fieldwork tributary continues in Collins's, (2013a; Ch. 17 and Appendix 4). Here will be found his futile and, perhaps, misguided, attempt to use his physics expertise to try to work out certain principles of gravitational wave data analysis, leading him to disagree with the physicists and become involved in another heated argument.

[14] This paper was probably written in 2003; unfortunately, the dates of drafts of the 2004 paper have been lost in the course of changes in word-processors and computer operating-systems.

This quotation takes the imitation game stream back to when Collins was running imitation games in the late 1980s and onward, for example, fruitlessly trying to prove that women would be better than men at detecting men pretending to be women but the notion that it was connected with interactional expertise was absent.[15] The idea that the imitation game could be used to test for the possession of interactional expertise is at least immanent around 2003:

> What I am saying is that it is possible to learn to say everything that can be said about bicycle-riding, car-driving or the use of a stick by a blind man, without ever having ridden a bike, driven a car, or been blind and used a stick. One could learn to pass the corresponding Turing Tests purely [from] spending enough time talking with the practitioners of the relevant domains without actually practising the practices. (Collins 2004b:127)

It was, however, not until 2004 that Collins and Evans began imitation game experiments with interactional expertise in mind. Our first experiments intended to explore the notion of interactional expertise were carried out on the colour blind and persons with perfect pitch (Collins et al. 2006). We then tested Collins's – and various others' -- ability to pass as gravitational wave physicists (Giles 2006). We then went on to test those who had been blind from an very early age and, following Evans's use of the game in a social research methods course, to look at sexuality and religion (Christians and non-Christians), we were awarded a large grant to look at such topics cross-nationally (Collins and Evans 2014).[16] The imitation game continues to be used by various groups on a large and small scale to look at many different kinds of cultural difference.

---

[15] There are two likely reasons for this failure. The first is that there is good reason to expect men and women to be quite good a pretending to be each other, so any difference is likely to be small and difficult to measure. The second is that, as our more recent experience testifies, the Imitation Game is much more complicated than it first appears and these early attempts represent a very 'noisy' and inefficient research design. Using the more sophisticated version of the Imitation Game developed over the past few years, it is possible to show that both male and female Judges can identify the pretender with a success rate that cannot simply be attributed to chance.

To repeat, the mainstream and the tributaries all originate from the sociological interpretation of Wittgenstein and that is why they all fit together. The philosophical stream disaggregates forms-of-life into practice and language components; the fieldwork tributary arises out of the focus of attention on the extent to which, in the absence of practice, the analyst has acquired the native form-of-life; the imitation game tributary is to do with seeing the Turing Test as a measure of how well an individual (or computer) can be embedded in a form-of-life under different circumstances.

### *4.1 Interactional expertise does not depend on the imitation game*

PK claim that the definition of IE underwent a change when we made passing the imitation game a criterion of possessing IE and Goddiksen says something similar. But we propose no such criterion. We say, as PK point out, the imitation game may 'set an impossibly high standard' for participants doing studies similar to Collins's studies of GW physics, where he passed an imitation game as a gravitational wave physicist after nearly a decade of deep immersion in the field. (Collins 2009:302, note 2, reinforces the point.)

Worse, it simply cannot be that we think that every person with IE must be able to pass the imitation game because most of our experimental results are statistical and compare the overall pass rates of different groups. Such aggregate results do not tell us about the capacities of individuals. The cases that test individuals are rare. Collins did pass the imitation game and this did show that he possessed IE in so far as the test was a reasonable one, but even here mathematical and algebraic questions were banned, personal questions were not asked (who was your PhD supervisor and what did he wear?), the dialogues had to be edited to remove obvious indicators of style, and the panel of judges had to concentrate on the substance of the answers. It is a feature of experiment that it is complicated and messy and this alone would make it impossible to claim that IE must be always be certified by passing an imitation game.

---

[16] The grant is an Advanced Research Grant awarded by the European Research Council (#269463, IMGAME, 2011-2016, €2.26M)

In spite of this, the imitation game is important in respect of the *linguistic* definition of IE and the strong interactional hypothesis; it is a way of empirically grounding this essentially philosophical definition even if the test does not have to be passed. Those who claim to have IE must be able to *think* about whether they could pass an imitation game in ideal circumstances. They may not be able to pass a real imitation game experiment for all the reasons that PK point out, not to mention that intensity and duration of fieldwork such as that of Collins is mostly impossible to achieve, but subjecting themselves to a thought experiment they should be ready to ask: 'In principle, could I pass an imitation game if questions and judgements were restricted to the technical substance of the domain, all indicators of previous diverse socialisation being removed?' The depth of understanding associated with answering 'yes' is indicated in the section of annotated dialogue between Barish and Collins quoted above. If the answer is 'no'– which, mostly, it will be – then the question switches to the areas of failure: 'What kind of question could I not answer and where would I succeed? Do I need more linguistic immersion with the group I am trying to understand or are my deficiencies not serious enough to vitiate the claims I am making about this group and their practices?'

## 5. The sociology and policy stream

As intimated, the new thought which underlies interactional expertise introduces a new way of thinking about how groups and individuals coordinate their actions in society. In many discussions interactional expertise is associated with scientific specialties but the more basic idea always had the potential to be applied more widely. The wider implications of the idea are shown in the sociology and policy column of Figure 1 (the last entry will be discussed in Part II). Shaded entries refer to policy, narrowly defined, whereas early unshaded entries refer to implications for sociology of science with the lower unshaded entries being implications for the working of societies as a whole. We start with the unshaded entries.

It was soon understood that a principal resource of managers of technical projects was interactional expertise (Collins and Sanders, 2007). It was very soon realised that peer review and the assessment of grant applications would be impossible without the reviewers having interactional expertise and this led to an understanding of its role in committee discussions and so forth. The concept also leads to new models of communication across linguistic divides (Collins, Evans and Gorman, 2007). These are implications for the sociology of science. The language and embodiment theme led to a better understanding of the role of sports coaches who are not themselves sportsmen and women, and so forth.

The concept also bears on fundamental questions about how societies function. Contributory experts were always recognised as possessing interactional expertise -- the relationship was transitive – but more recently, it was realised that the larger part of all, or nearly all, expertise is interactional expertise, since it is fluency in a high level practice language that provides the framework and direction for the acquisition of contributory expertise. This explains how specialists who, by definition, do not do each others' jobs, can coordinate their practical actions in a 'complex' division of labour where by 'complex' division of labour we mean that successful coordination depends on understanding of others' worlds of practice.[17] Since all contributory experts are interactional experts too, we now think that those few who gain their practical understanding through the language alone – such as the social scientists and high level journalists we first had in mind – should be referred to as 'special interactional experts'.

## *5.1 Ubiquitous interactional expertise*

We now want to take the opportunity to extend the application of the concept of IE still further into the understanding of how societies work. Not just groups of specialists, but large parts of society as whole are held together by interactional expertise! For example, we think that this is how it is that, say, men can understand women and women can understand men well enough to coordinate their activities in ordinary social life in spite of the fact that the coordination often depends on mutual understanding of practices which they cannot all experience directly. We claim that this is how every social group that is distinct in terms of its practices interacts with other social groups when that interaction is 'complex' – ie, when that interaction demands understanding of and coordination with the other's practical life. This does not mean that, for example, in Western societies, every man understands every woman or every woman understands every man nor that in the case of those who do understand each other the understanding is complete. Nevertheless, without a concept like interactional expertise we could not understand how it is that the lives of men and women, *per se*, are not much more closed to each other and it would be hard to understand how their actions could be coordinated beyond their following strict sets of rules for interaction – non-complex interaction via mimeomorphic actions – as may be the case in societies where men's and women's lives are relatively separate. The question of how far and how deeply the understanding stretches is exactly the kind of question that can be approached with large scale imitation games.

---

[17] In complex division of labour it is polimorphic actions that must be coordinated (Collins and Kusch 1998); simple division of labour works with mimeomorphic actions (at least, in theory). To make the idea apply we also need the 'fractal model'. This and other ideas are explained in Collins, 2011.

Interactional expertise has, then, the potential to be a basis for a theory of society. Sometimes actions are coordinated, to a greater or lesser degree, through the shared practices and language of different groups; sometimes they are coordinated, to a greater or lesser degree, through shared language without shared practices; sometimes they are coordinated, in a more shallow way, through practices and language which are very widely distributed across whole societies and go only a little way to explain the interactions of smaller groups; sometimes they are coordinated without common understanding, through mimeomorphic actions alone – or we should say that sometimes attempts are made to bring about such coordination though its successful implementation is much rarer than used to be thought.[18]

Society, then, is held together in a significant part by language.[19] With this development IE is in danger of becoming coextensive with language as a whole but it still retains meaning when it is remembered that the point of the concept is that it delivers practical understandings of the other without requiring that the practices of the other be practised. Interactional expertise is going on only when that is happening. So, not all language is interactional expertise but there is a huge amount more interactional expertise about than we thought when the concept was invented.

---

[18] It would be interesting to compare this theory with Emile Durkheim's (1893 / (2013)) idea of division of labour. An important difference seems to be that, as a result of analysis of scientific knowledge which arises out of the sociological interpretation of Wittgenstein, Durkheim's notion of organic solidarity is seen as much more deeply invested with mechanical solidarity – that is, we now know that social backgrounds must be shared before labour can be divided in any complicated way. It is this sharing that is accomplished by interactional expertise. It is interesting to reflect on how exactly opposite this approach is to that of Latour, who wishes to replace Durkheim, not with a deeper understanding of culture but with Gabriel Tarde's notion of association (see e.g. Latour 2005).

[19] See also Collins (2011) for a discussion about how human society is different to animal society in this regard: '... specialities are distributed among dogs: there are pointers, foxhounds, chihuahuas, and so forth. There is, however, nothing that can link the differing experiences of these dog specialists into a joint domain of dog practice. The pointer is unable to come to know anything of the foxhound's world per foxhound; the foxhound is unable to come to know the world of the chihuahua. The world of the dog cannot combine the embodied specialities of all dogs because there are no doggy practice languages – practical activities among dogs cannot be glued together by language.'

We do not want to call large social groups such as men and women 'special interactional experts' so we introduce the term, 'ubiquitous interactional expertise'. Just where special interactional expertise turns into ubiquitous interactional expertise is not clear (how should we describe the blind or the religious?) but this is simply a matter of comfortable usage.

To sum up, the term interactional expertise now has three references

1) **Interactional expertise**: This is the ability to talk a practice language and includes contributory experts as well as others
2) **Special interactional expertise**: This applies only to those who have no contributory expertise in certain domains with which they nevertheless make a particular effort to engage. These include sociologists, high level journalists, the disabled and non-player sports coaches who learn a language in order to facilitate interaction between language communities. Whether 'special' applies to the interactional expertise of managers of technological projects, or whether they are better seen as contributory experts using referred expertise, is unclear but it does not seem to matter very much. We discuss the issue at greater length in Part II.
3) **Ubiquitous interactional expertise**: This is like the above but refers to members of the general population where the practice language is at a high level such as being a man/woman or belonging to any other identifiable social group which others encounter routinely and come to understand without sharing the practices

The second two kinds of interactional expertise will come in various strengths depending on the degree of linguistic socialisation that has been achieved.

## *5.2 Policy and the Third Wave*

In the 2002 Third Wave paper IE is given its first small role under that name. Thus, in the case of Wynne's sheep farmers' it is their contributory not their interactional expertise that is said to justify their contribution:

> To produce the optimum outcome, the [qualified] scientists needed to have the *interactional expertise* [in sheep farming] to absorb the expertise of the farmers. (Collins and Evans 2002:256)

IE was presented in the Third Wave paper as an element in a theory of expertise that was meant to appeal widely to social scientists. The first published definition of the term is:

> *interactional expertise*: This means enough expertise to interact interestingly with participants and carry out a sociological analysis. (Collins and Evans 2002:254)

Because certain of our critics want to make this the main definition of IE it is important to note that there is no further elaboration on the meaning of IE in the paper and this is because what we were trying to do in the 2002 paper was establish 'Wave 3' of science studies by bringing the concept of expertise to centre stage in place of the concept of truth. We knew that we had an uphill task and at the time of writing we felt we needed to show what a theory of expertise might look like:

> To show that our argument is more than a programmatic gesture, we will indicate one way to start to build a normative theory of expertise, and what it would mean for technical decision-making. There are, no doubt, many other ways to go about such an exercise, but to focus attention on the goal by providing an example of one approach to it is at least a start. (Collins and Evans 2002:238)

A couple of years later it was again emphasised:

> The need for an in-between category [of expertise] originally arose out of a set of concerns distant from those discussed here, hence the name 'interactional expertise' (it might have been something like 'linguistic culture'). The chief concern was to persuade sociologists of a 'relativist' or 'social constructivist' bent to take expertise seriously – to treat it as something *real*. The rhetorical trick was to persuade such sociologists' to reflect upon their own expertises. (Collins 2004b:127)

Thus the first definition of IE focuses on the expertise required to do *sociological analysis* of science, not the expertise required to do science.

The first extensive discussion of the potential policy implications of interactional expertise is found in the, 2004, *Gravity's Shadow* following the debate with Barish about noise mentioned above. There it is argued that his possession of IE does *not* justify technical inputs in this esoteric science, but for sociological not epistemological reasons:

> The difference [between Barish and me] is a difference in the forms of life in which we are embedded, he as scientist, me as analyst. In nearly every case (we will look for an exception at the end of the chapter), interactional expertise is not a sufficient basis to make judgments of this sort [the level of noise in the interferometer]. Thus, irrespective of any fleabites I might inflict, Barish's scientific judgment—the judgment he will act on—remains that there are no unknown fundamental noise sources to be found in LIGO, and he has to take responsibility for the judgment. In spite of (what we will allow to be, for argument's sake) my fairly creditable performance in the argument, at this level I am not *contradicting* him, because nothing turns on what I think. ... The scientists have already discounted their uncertainties even though they may not have articulated them. Like any other decision makers, they live with the chance of being wrong but must still act. This is the expertise of the core set, and this is where decision-making rights in matters of esoteric science should remain. If this book makes any difference to esoteric science, it should affect only the way scientific judgment is thought about, not the way it is done. (Collins 2004a:778–9)

A key modification about 'decision-making rights' is, however, developed in the 'possible exception', which imagines a more policy-relevant scenario – a case where we:

use our imaginations to convert gravitational wave science into a public domain science. Imagine ... that immersion in a certain (largish) flux of gravitational radiation, combined with living near an overhead electrical power line, caused cancer. ... In such a circumstance, who should be involved in decision making about whether [certain detection claims should] be taken seriously? What kinds of levels of expertise and political sensitivities should be brought to the table in these circumstances? ... The political decisions have to stand on a foundation of science. Yet science cannot work at the pace of political decision making. ... the classification of types of scientific expertise, including contributory expertise, interactional expertise, and referred expertise, might be useful in the discussion ... I suspect that in these circumstances the relative value of interactional expertise might increase as compared with its weight within core science. ... I offer the idea for debate. (Collins 2004a:781–2)

It is that debate that is continuing on these pages.

What is assumed in the quoted passage is that the social analyst with IE would be exposing the residual interpretative flexibility in the findings in the face of core scientists' tendency to gloss over uncertainties so as to reach conclusions upon which decisions about actions can be taken. This is not to say that scientists themselves could not work out that if there were serious public health consequences any residual uncertainties should be given more salience but, for argument's sake, let us imagine this to be the prerogative of interactional experts or that the interactional experts had a range of additional public interest viewpoints not salient to the core scientists. It seems that Collins is arguing here that under these circumstances special interactional experts could, reasonably, have greater influence on the strength of the technical conclusions than in the case of a purely esoteric science; this seems right.[20]

But, we want to stress that it still has to be IE that is brought to bear, not just additional public interest viewpoints in the absence of IE. In contrast, it was the unconditional importance of the public interest viewpoints on the content of science *even in the absence of IE* or any other relevant contributory expertise that was being argued for by mainstream STS in the early 2000s under the label of 'lay expertise'. It is this argument that we were resisting by trying to resurrect the importance of expertise. We agreed that Wave 2 had shown how influenced science was by politics but we were arguing that this influence was intrinsic and should never be extrinsic.[21] We were not, and are not against the importance of additional public interest

---

[20] And to this extent, meshes well with PK's; they seem to specially interested in cases where stressing the uncertainties of science are the prerogative of interactional experts or that the interactional experts have a range of additional public interest viewpoints not salient to the core scientists.

[21] These terms, introduced in the 2002 paper, imply that while the second wave of science studies had shown that science is intrinsically political, scientific findings cannot be justified by their political implications and, it

viewpoints affecting policy conclusions but under most circumstances we argue that their domain is the political phase of such arguments, not the technical phase.[22] To apply the point to the example discussed in *Gravity's Shadow*, even if the cost of being wrong about non-existence of high fluxes gravitational waves was an avoidable increase in the incidence of cancer we should not expect the core-group of gravitational wave scientists to change their minds about the detection of high fluxes. Instead, what might change, conceivably triggered by the remarks of persons with IE, would be the way they would present the levels of uncertainty to whoever was making the decisions in the political sphere, advising them that if it was purely a matter of the future of gravitational wave physics the level of uncertainty was too low to justify further research into the detection of high fluxes of gravitational waves but, if it was a matter of potential loss of life, then the balance of risk would be different and a more cautious approach could be justified.

## 6. Recent critiques of interactional expertise

With this re-visit to the idea of interactional expertise in mind we can look at the recent critiques and proposed amendments of IE by Plaisance and Kennedy (2014) and by Goddiksen (2014), along with the reply to the later by RGD. PK want to go back to the definition found in the 2002, Third Wave, publication. To repeat, this was 'enough expertise to interact interestingly with participants and carry out a sociological analysis'. But this, even if it were enough IE to justify a technical contribution, which it is not, is not really in the

---

follows, that any time scientists become aware of political influence on their work they should strive to eliminate it.

[22] The distinction between technical and political phase, which is also introduced in the 2002 'Third Wave' paper, is further discussed in papers by Collins, Weinel and Evans (2010, 2011); it is also a central element of PartII.

spirit of IE in so far as the meaning of the idea can be extracted from the story of its origins; it is not even a very careful expression of the fieldwork tributary because it was intended for another purpose: to try to persuade sociologists to take expertise seriously. Of course, PK and Goddiksen have every right to try to change the idea of IE in directions not envisaged by its authors but we should be clear that what they propose is transformation not evolution. And this is not surprising since they, at least in the case of PK, are not drawing on the same conceptual tributaries as IE but from a desire to further democratise science.

> By helping to identify those with potentially relevant experience and expertise, and legitimizing their knowledge, the use of this concept [IE] can lead to scientific research that recognizes a more diverse set of epistemic resources ... [and help] contributory experts acknowledge the potential contributions of lay communities and other stakeholders p 68

Pursuing this point they demand an alternative 'pluralistic' concept of IE, one resting on "a more diverse set of expertise that can lead to better scientific knowledge, as well as more democratic technical decision-making" (p61) and which uses "the original, more holistic definition of IE as 'interacting interestingly' with contributory experts", "[s]hifting towards a broader definition of interactional expert – one who has the linguistic ability to interact interestingly and productively with contributory experts"

We, in contrast, are not driven by a desire to democratise science further but to improve technological decision-making in the public domain – to democratise it to just the right amount, as one might say: sometimes this will mean more participation by more diverse social groups and sometimes it will mean excluding social groups who wish to participant as experts but have no legitimate basis on which to make this claim – sometimes 'more democracy' and sometimes 'less democracy', or, rather, less populism. Even when wider participation is the right solution, IE is only occasionally the appropriate route. For example, IE plays an important part (as we see it) in the success of the AIDS treatment activists described by Epstein (Epstein 1996) and in our claim (Collins and Evans, 2007) that, if the idea of IE became established, it could help with the credibility of sociologists of forensic scientists in courtrooms and in the analysis of doctor-patient relationships and so on.[23] We will discuss additional possibilities below. But on other occasions the false impression that a person understands an issue to do with science and technology leads in the wrong direction and there is *too much* participation – that is, the problem of extension is exemplified. Even in the 2002 paper invoked by PK, we offer examples where there are too many witnesses who do not understand how to interpret the spectacles of a train crashing into a nuclear fuel flask without damaging it and of a plane fuelled with anti-misting kerosene bursting into flames as it hit the ground. In these cases the public interest might have been better served if the interpretation had been left to a narrow group of experts. This means that, as one might say, less democracy would have been better!

Another key example of where we need less inclusion is anti-vaccination campaigns where members of the public, reinforced by journalists (and, sadly, even some social scientists), come to believe they have a legitimate contribution to make to medical science when they do not. The idea of interactional expertise should never be responsible for transforming a parent's heartfelt and vivid experience of their child developing symptoms of autism shortly after the administration of MMR vaccine into the belief that they have a legitimate claim that the vaccine caused the symptoms.  In spite of the huge stake parents held in the vaccination business and the immediacy of their experience – something that clearly engaged the newspapers, the public and some social scientists – IE and the Third Wave must indicate to parents: 'tragic though your experience is, being a parent does not provide enough understanding of the technical discourse for your stake and your experience to count against the opinion of the technical community when it comes to establishing a causal connection between MMR and autism.'

---

[23] Epstein (2011) considers that his respondents' IE was less important to their success than we make out. The medical use is discussed in Evans and Crocker (2013) and Wehrens (2014)

PK seem to grasp the point when they say:

> 'to ensure that experts aren't bombarded with outside perspectives IEs must know the key concepts, be able to speak the language, and understand the expert community' (p. 66)[24]

But they do not tell us how we and the rest of the population are to know whether they 'speak the language and understand the expert community'.  It is a characteristic of many, not least anti-vaccination campaigners, and their constituency, that they believe they know these things when they do not.  If we are to avoid the dangers that the Third Wave set out to address, at the very least we need a clear, non-circular, definition of IE, where 'non-circular' means that having a strong and engaging view about the science in question does not automatically lead to the conclusion the person also possesses IE.  The MMR parents and journalists had a strong and engaging view about vaccination, as well as the largest possible stake in the matter, but they did not possess IE in the matter of the safety of MMR vaccination and therefore had no legitimate contribution to make to technical debates about vaccine safety. We need clearly defined categories of expertise if they are to hold the line in this kind of controversy.[25]

---

[24] Incidentally, the sheep farmers did not meet this criterion!!

### 6.1  Partial interactional expertise and the slippery slope

Holding the line is made more difficult because not all IE is the maximal interactional expertise postulated by the strong interactional hypothesis. There will be degrees of interactional expertise. Thus we could easily find ourselves on a slippery slope with imaginary imitation games and the like. How, else might one judge the level of IE possessed by oneself or by another?

The definition of IE must have to do with acquiring language and understanding, and thus tacit knowledge, through interaction with the expert community. What the criterion must not become is membership of some protest group supported by the popular literature and the press, which PK are in danger of making it. Nor must it be normal classroom education to which it slips under Goddiksen's (2014) treatment. One possible criterion is, then, to consider the length of time that the person has spent immersed in the spoken discourse of the target community. Persons who are contributory experts are, of course, maximal interactional experts but there is no question to answer in that case.[26] Where there is a question to answer one can ask, is the immersion deep and lifelong, as in the case of the blind, or Madeleine? Is the immersion lifelong but not as deep, as in the case of the relationship of men and women in liberal societies and other social groups who share a common natural language and partially share their practice languages? Is it a good proportion of a life as in the case of the chronically ill? Is it the ten years of fairly intense exposure of a fieldworker like Collins? Is it the one or two years of more regular fieldworkers? Is it some other kind of interaction with the oral community of experts driven by interests outside of scientific research? The answer to that question should be clearly stated.

---

[25] Having IE does not necessarily guarantee decision-making rights either, as will be discussed in more detail in Part II.

[26] In practice, the extent to which interactional expertise is *manifested* by contributory experts will depend on their reflective and interactive abilities. Where these are poor, then the interactional expertise will be latent (see e.g. Collins and Evans 2007:36–40)

Another way to establish a floor to the concept – to say what interactional expertise is not – is to adopt a Wittgensteinian tactic. Think about uses not meanings and take some paradigm cases of use as a way of setting lower limits.

To find uses that can flow back toward the definition of a floor, we suggest two cases. The first is the invisibility within the gravitational wave community of Joseph Weber's, 1996, published paper on the correlation between his claimed early gravitational wave sightings and gamma ray bursts (Weber and Radak 1996). The account of the incident is first set out on pps 366-8 of *Gravity's Shadow*, but has been widely referred to since. The point is that to possess interactional expertise in a domain means being, or having been, sufficiently deeply immersed in the oral community of specialists to know how the community assesses the credibility of the range of publications that bear on a topic. There are always more publications than credible publications and before one can even decide how to treat the ones that are not credible one has to be able to recognise which are which in the eyes of the core community.[27] The second paradigm case is the MMR vaccination revolt: no definition of technical expertise must allow the typical MMR parent to justify their vaccination choice on the basis of their expertise in the relationship of MMR and autism in spite of their deep engagement with the problem.[28] In the absence of deep engagement with the oral community, the MMR parents would not be able to recognise when a scientific paper, written with all the persuasiveness of science's literary technology, was to be taken into account and when it should be ignored as cranky or 'past its sell-by date'.[29] Goddiksen's approach to IE cuts it off at too low a level to correspond to either of these cases and PK's usage is also in danger of sinking it through the floor.

---

[27] RGD discuss the principle of this point without the example.

If the definition is not kept reasonably tight and the question about 'how much IE does this person or group possess' is not asked then the danger alluded to in the introduction – the invention of 'interactional expertise lite' – arises. IE lite could turn the Third Wave project on its head by justifying all kinds of technical intervention by those who are not technical experts according to the Periodic Table of Expertises. IE lite could come to serve the popular call for unlimited democratisation of science to which the term 'problem of extension' refers. IE lite, like other 'lite' terms, might do well in the citation indexes but it would be a shame to lose the deep philosophical point and return to the baggy old idea of 'lay expertise' which IE, and the Third Wave of science studies as a whole, set out to eliminate. The tendency to convert 'real IE' into IE lite would be exacerbated were Goddiksen's (2014) interpretations to be accepted. As RGD point out, Goddiksen misses the tacit knowledge/socialisation dimension of the concept when she tries to replace what can only be acquired through deep immersion within the oral community of research scientists with what is learned in the classroom. The fractal model (Collins 2011), which Goddiksen understands well, means that we have to say that teaching assistants possess some expertise but if it is IE, and even CE, it is IE and CE, in 'science-as-taught-in-classrooms' not in frontier research science. This classroom exposure is far from immersion in the social world of frontier science. We think Goddiksen and Plaisance and Kennedy may have missed the point of 'real IE' because they do not share the philosophical tradition from which it arose.

---

[28] We believe that episodes such as the revolt against MMR vaccine are something that social scientists should not be endorsing but criticising. Of course, not all social scientists take this view. For instance, Wynne argues that the parents were right to demand more research on the relationship, ignoring the problem of scarcity and the fact there was no more evidence that MMR caused autism than that anything else causes it. See Collins (2014), for an attempt to use the Periodic Table of Expertises as a way of extending the legitimate domain of expertise without justifying such things as the MMR revolt

[29] See e.g. (Collins and Weinel 2011; Weinel 2007)

We agree with PK that *some* scientific and technological decisions should involve more non-certified persons than they often do. We agree, further, that a subset of these cases would be progressed by the recognition of the interactional expertise of non-certified, non-core-set persons. These cases are the exception, however. More often, citizens' contributions to science have nothing to do with their IE in the scientific domain but are a matter of bringing a new *contributory* expertise to the table. Furthermore, possessors of IE ought sometimes to be excluded from making a contribution because their contributory expertise is outdated, as with Nobel Laureates' Duesberg and Mullins' unfortunate contribution to Thabo Mbeki's decision not distribute anti-retrovirals to pregnant South African women (Weinel,2010). Sometimes, then, more democratisation is to do with possessing IE, sometimes it is recognising new contributory expertises, and sometimes it is other things that will be discussed in Part II. It seems to us that PK should be making the case they want to make not from IE but from what can be offered by the overall programme called Studies of Expertise and Experience (SEE). Their focus is on who can contribute to a science-based policy issue rather than who can talk fluently about the underlying science. It is to this question that we will turn in Part II.

For now we conclude that those who do want to legitimate their contributions by claiming to have IE in some technical domain – a very small set of cases among those that involve citizens making contributions to technological debates – can set a floor to the concept by thinking about the following:

1) imitation game
   (a) Can an imitation game be passed and, if not, what are the typical failures and do these vitiate the work on which possession of IE was based?
   (b) Failing this, and very much a second-best, could an imaginary imitation game conducted under ideal circumstances be passed and, if not, what would the typical failures comprise and would these vitiate the work on which possession of IE was based?
2) How much time has the person who is said to have IE spent embedded in the linguistic discourse of the core-group of scientists in the area?
3) Does the definition of IE that the researcher or activists have in mind always lead to more inclusion and participation or are their cases where it leads to allows non-experts to be identified and excluded from technical decision-making; in particular, how would the definition work in the case of the MMR vaccination controversy?
4) Does the putative possessor of IE understand, without formal explanation, which are credible and which are non-credible publications in the technical domain in question?

## 8. Acknowledgements

In part, the idea of interactional expertise grew out of research supported by the UK Economic and Social Research Council, notably three grants to Harry Collins: ESRC (RES-000-22-2384) £48,698 `The Sociology of Discovery' (2007-2009); ESRC (R000239414) £177,718 `Founding a New Astronomy' (2002-2006); ESRC (R000236826) £140,000 `Physics in Transition' (1996-2001)